\documentclass{nature}
\usepackage{amsmath,amssymb,amsfonts}%
\usepackage{lineno}
\usepackage{stmaryrd}
\usepackage{bbm}
\usepackage{mathrsfs}
\usepackage{tipa}
\usepackage{graphicx}
\usepackage{dcolumn}
\usepackage{epstopdf}
\usepackage{microtype}
\usepackage[colorlinks,linkcolor=blue,urlcolor=blue,citecolor=blue]{hyperref}
\usepackage{multirow}
\usepackage{url}
\usepackage{upgreek}
\usepackage[utf8]{inputenc}
\usepackage[english]{babel}
\usepackage{xcolor}
\usepackage[normalem]{ulem}
\usepackage[title]{appendix}%
\usepackage{csquotes}
\MakeOuterQuote{"}
\usepackage{physics}
\usepackage{array}
\usepackage{times}
\usepackage[numbers,sort&compress]{natbib}
\usepackage{graphicx}
\makeatletter
\let\saved@includegraphics\includegraphics
\AtBeginDocument{\let\includegraphics\saved@includegraphics}
\renewenvironment*{figure}{\@float{figure}}{\end@float}
\makeatother

\newcommand{\hpstar}{Center for High Pressure Science and Technology Advanced Research, Beijing 100193, China}
\newcommand{\tsung}{Tsung-Dao Lee Institute, Department of Physics and Astronomy, Shanghai Jiao Tong University, Shanghai 200240, China}
\newcommand{\bayarea}{Quantum Science Center of Guangdong-Hong Kong-Macao Greater Bay Area, Shenzhen, 518045, China}
\newcommand{\sust}{Department of Physics and Guangdong Basic Research Center of Excellence for Quantum Science, Southern University of Science and Technology, Shenzhen, 518055, China}
\newcommand{\pku}{International Center for Quantum Materials, School of Physics, Peking University, Beijing 100871, China}
\newcommand{\baqis}{Beijing Academy of Quantum Information Sciences, Beijing 100913, China}
\newcommand{\iop}{Beijing National Laboratory for Condensed Matter Physics, Institute of Physics, Chinese Academy of Sciences, Beijing, 100190, China}
\newcommand{\tsing}{Department of Physics, Tsinghua University, Beijing, 100084, China}
\title{Multimodal Terahertz Spectroscopy of the Pairing Symmetry and Normal-State Pseudogap in (La,Pr)$_3$Ni$_2$O$_7$ Films }

\begin{document}
\maketitle

\author{Shuxiang Xu$^{1, 2,\text{\#}}$, Guangdi Zhou$^{3, 4,\text{\#}}$, Hao Wang$^{2,\text{\#}}$, Tianyi Wu$^{2}$, Wei Wang$^{3,4}$, Liyu Shi$^{2}$, Dong Wu$^{5}$, Haoliang Huang$^{3, 4}$, Xinbo Wang$^{6,\dagger}$, Jinfeng Jia$^{3,4,7}$, Qi-Kun Xue$^{3,4,8,\dagger}$, Zhuoyu Chen$^{3,4,\dagger}$, Tao Dong$^{7,2,\dagger}$, Nanlin Wang$^{7,2,5,\ddag}$}

\begin{affiliations}
	
	\item \hpstar 
	\item \pku 
	\item \bayarea 
	\item \sust 
	\item \baqis 
	\item \iop 
    \item \tsung 
	\item \tsing 
	 
	 $^{\text{\#}}$These authors contributed equally to this work
	  
	 $^{\dagger,\ddag}$Corresponding authors, E-mail: xinbowang@iphy.ac.cn, xueqk@sustech.edu.cn,
	 chenzhuoyu@sustech.edu.cn, taodong@sjtu.edu.cn, nlwang@sjtu.edu.cn
	 
\end{affiliations}

\clearpage

\setlength{\parskip}{4pt}

\begin{abstract}
The discovery of ambient-pressure superconductivity in compressively strained (La,Pr)$_3$Ni$_2$O$_7$ thin films has intensified efforts to identify the pairing mechanism. However, the symmetry of the superconducting order parameter and the character of the normal state remain unsettled. Here we combine bulk-sensitive terahertz (THz) time-domain spectroscopy with THz third-harmonic generation to present spectroscopic insights into these issues. Linear THz spectroscopy reveals a bulk superconducting response in the  (La,Pr)$_3$Ni$_2$O$_7$ films, evidenced by the suppression of low-frequency spectral weight below the onset critical temperature, $T_\mathrm{c}^{\mathrm{onset}}$. A weak coherence peak near $T_\mathrm{c}^{\mathrm{onset}}$, together with substantial residual low-frequency conductivity as $T\to 0$, is consistent with disordered $s_{\pm}$-wave pairing. In the nonlinear regime, the third-harmonic signal rises sharply on cooling through $T_\mathrm{c}^{\mathrm{onset}}$, providing an independent signature of the transition. Strikingly, the nonlinear response persists above $T_\mathrm{c}^{\mathrm{onset}}$, pointing to either disorder-enhanced nonlinearity or a distinct correlated normal state. Motivated by angle-resolved photoemission spectroscopy on similarly grown films that identifies a comparable temperature scale, we associate the anomalous normal-state terahertz nonlinearity with a pseudogap. These results establish (La,Pr)$_3$Ni$_2$O$_7$ as a bulk superconductor with $s_{\pm}$-like pairing that coexists with, and may compete with, a distinct ordered state, providing a platform for exploring unconventional superconductivity beyond cuprates and pnictides.

\end{abstract}

\maketitle 
\clearpage

\vspace{18pt}
\noindent {\large \textbf{Introduction}}

The recent discoveries in nickel-based Ruddlesden–Popper (RP) superconductors has revitalized the search for unconventional pairing mechanisms \cite{Sun2023,Hou_2023,Zhang2024high,Wang2024,Zhu_2024,zhang2024,Liq_2024,Shi2025}. In contrast to the largely single-orbital cuprates, RP nickelates host a multiorbital electronic structure near the Fermi level and typically require external pressure to stabilize superconductivity, complicating the identification of the dominant pairing interaction \cite{Yang2023,Luchen2024,Fan2024}. Moreover, in cuprate and iron-pnictide high-$T_\mathrm{c}$ superconductors, pseudogap, strange-metal, nematic, and charge- or spin-ordered phases are recurrent features of the correlated phase diagram \cite{Keimer2015,Si2016}. Whether a comparable landscape of intertwined orders is realized in nickelates—and how any such orders relate to superconductivity—remains an open question and a key point of comparison across high-$T_\mathrm{c}$ families.

A defining feature of superconductivity is  the single‑particle excitation gap $2\Delta$, whose magnitude, symmetry, and momentum dependence are intimately linked to the underlying pairing mechanism. Elucidating the symmetry and structure of this gap is thus a key step toward understanding superconductivity in nickelates. The recent stabilization of ambient-pressure (La,Pr)$_3$Ni$_2$O$_7$ thin films has enabled detailed experimental characterization \cite{Ko2024,Zhou2025,Hao2025}, primarily through surface-sensitive angle-resolved photoemission spectroscopy (ARPES) and scanning tunneling microscopy (STM) \cite{shen2025LPSNO,10.1093/nsr/nwae429,wang2025,wangbai2025,fan2025}. However, bulk‑sensitive spectroscopic measurements, which are crucial for establishing intrinsic superconducting properties, remains scarce. Time‑domain terahertz spectroscopy provides a powerful bulk‑sensitive probe of superconductivity, directly accessing the low‑energy optical conductivity and thereby constraining the gap structure and pairing symmetry \cite{PhysRevB.50.6307,Bilbro2011,Cheng2024}. Another hallmark of superconductivity is its collective excitations, which encodes the fingerprints of the order parameter. Following the seminal observation of the Higgs mode in conventional NbN superconductors \cite{Matsunaga2013,Matsunaga2014}, nonlinear THz spectroscopy has emerged as a sensitive tool for identifying superconductivity and for probing its coupling to nearby symmetry‑broken phases \cite{Katsumi2018,Chu2020,Yuan2024,Kim2024} or to disorder \cite{wang2025anomal}. Moreover, nonlinear THz responses can interrogate the normal state from which superconductivity emerges, offering additional insight into the correlated electronic background and potential competing orders \cite{Katsumi2018,Chu2020,Feng2023,Yuan2024}.

Here we combine linear and nonlinear THz spectroscopy to probe the superconducting (La,Pr)$_3$Ni$_2$O$_7$ thin films grown on SrLaAlO$_4$ substrates. Linear THz conductivity reveals a superconducting gap opening below $T_\mathrm{c}^{\mathrm{onset}}$ alongside an unusually large residual quasiparticle response, and yields a large London penetration depth, $\lambda_\mathrm{L}\simeq 620$ nm at 2 K, implying a strongly reduced superfluid density compared with cuprates and iron-based superconductors \cite{Boovi2016,Mller1990,Liprl2008,Williams2010}. Together with a weak coherence peak below $T_\mathrm{c}^{\mathrm{onset}}$, these signatures are consistent with sign-changing $s_{\pm}$ pairing in the presence of strong disorder. In the nonlinear response, the third‑harmonic generation (THG) signal rises sharply below $T_\mathrm{c}^{\mathrm{onset}}$ but persists into the normal state and exhibits a reproducible kink near 100 K, indicating either disorder‑enhanced nonlinearity or a distinct pseudogap phase. By comparison with the characteristic temperature identified by ARPES on similarly grown films \cite{shen2025LPSNO}, we associate the anomalous normal‑state THz nonlinearity with a pseudogap. Together, these results indicate that the entire film thickness of (La,Pr)$_3$Ni$_2$O$_7$ enters a phase-coherent superconductivity with disordered, sign‑changing $s_{\pm}$ pairing and reveal an additional ordered state emerging above $T_\mathrm{c}^{\mathrm{onset}}$.

\vspace{18pt}
\noindent {\large \textbf{Spectroscopic evidence for a disordered $s_{\pm}$-wave like superconductivity}}

\begin{figure}
	\includegraphics[width=1\columnwidth]{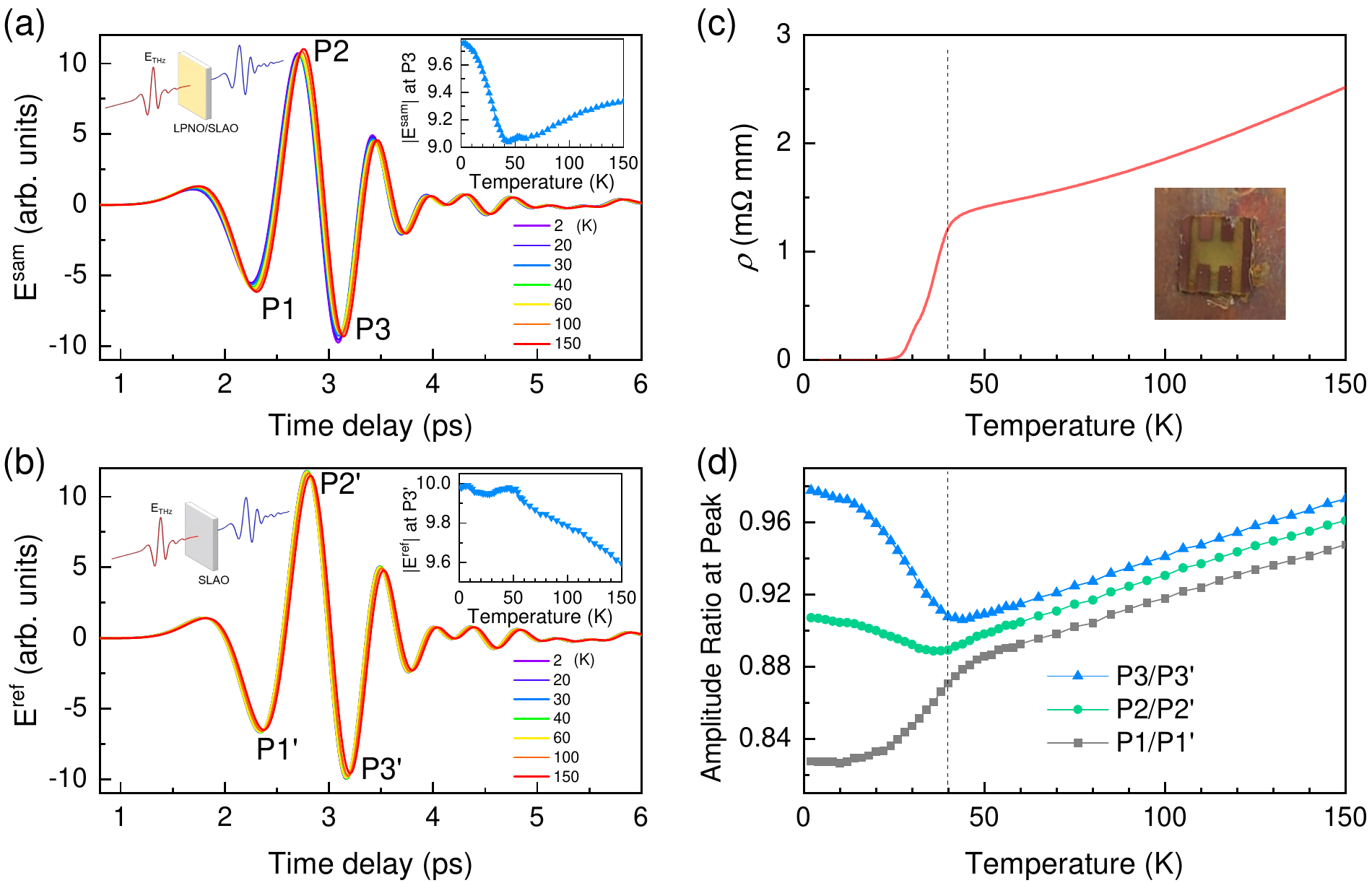}
	\caption{\textbf{The fingerprint of superconductivity observed in the linear THz response of (La,Pr)$_3$Ni$_2$O$_{7}$ films.
		} (a) Terahertz time domain waveforms transmitted through LPNO/SLAO sample at selected temperatures. The three main peaks are labeled P1, P2 and P3, respectively. The insets show the experimental schematics (top left) and temperature-dependent field $| E ^{\text{sam}}|$ at the position of peak P3 (top right). $| E ^{\text{sam}}(\text{P}3)|$ decreases rapidly below 40 K and then increases upon warming. (b) Terahertz time domain waveforms transmitted through the SrLaAlO$ _{4} $ substrate at selected temperatures. The corresponding peaks are labeled P1$ ^{'} $, P2$ ^{'} $ and P3$ ^{'} $, respectively. The insets show the experimental schematics (top left) and the temperature dependence of $ | E ^{\text{ref}}|$ at peak P3$ ^{'} $  (top right). $| E ^{\text{ref}}(\text{P}3^{'}) |$ remains constant below 50 K and then decreases with increasing temperature. (c) Temperature-dependent resistivity of (La,Pr)$_3$Ni$_2$O$_{7}$ film. The inset displays an optical image of LPNO/SLAO sample with platinum electrodes. (d) Temperature evolution of the peak amplitude ratio Pi/Pi$ ^{'} $ (for i = 1, 2, 3). The black dashed lines in (c) and (d) mark the onset of superconductivity transition temperature $T_\text{c}^{\text{onset}}\simeq $ 40 K. 
	} 
	
	\label{Figure1}
\end{figure}

We investigated 7‑nm‑thick (approximately three unit cells) (La,Pr)$_3$Ni$_2$O$_7$ films grown on SrLaAlO$_4$ substrates (LPNO/SLAO). Transport measurements confirmed superconductivity in all films before optical characterization. Figure \ref{Figure1} a,b shows the time-domain THz waveforms transmitted through the batch \#1 LPNO/SLAO sample, $E_{\mathrm{sam}}(t)$, and the bare substrate, $E_{\mathrm{ref}}(t)$, at various temperatures. Although the atomic thickness and low carrier density of the films render temperature‑dependent changes in $E_{\mathrm{sam}}(t)$ subtle, the sample and reference exhibit distinct behavior. Specifically, the amplitude of $E_{\mathrm{sam}}(t)$ shows a non‑monotonic anomaly near 40 K (inset of Fig. 1a), whereas $E_{\mathrm{ref}}(t)$ remains unchanged below 50 K (inset of Fig.\ref{Figure1} b). This contrast demonstrates that the spectral changes below 40 K originate from the intrinsic response of the (La,Pr)$_3$Ni$_2$O$_7$ layer. Consistently, the temperature dependence of the peak‑field ratios $P_i/P_i'$ ($i=1,2,3$; Fig. \ref{Figure1} d) exhibits a pronounced inflection near 40 K: upon cooling, $P_1/P_1'$ and $P_2/P_2'$ decrease, whereas $P_3/P_3'$ increases. This inflection coincides with the superconducting transition identified by transport measurements on the same film (Fig. \ref{Figure1} c).

\begin{figure}
	\includegraphics[width=1\columnwidth]{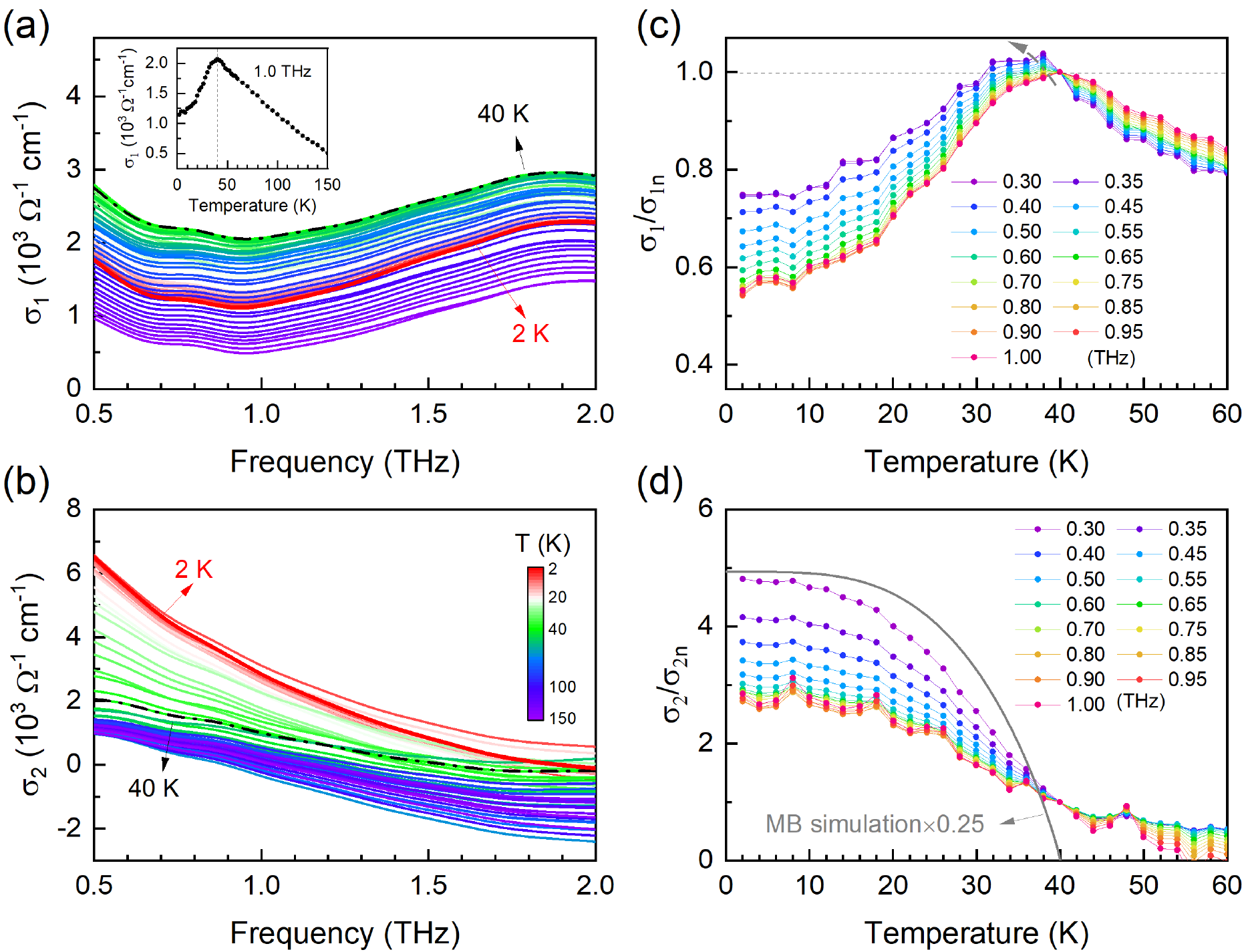}
		\caption{\textbf{Disordered $s_\pm$-wave superconductivity evidenced by the THz conductivity of (La,Pr)$_3$Ni$_2$O$_{7}$ films.} (a, b) Frequency-dependent (a) real part $ \sigma _{1}$ ($\omega$) and (b) imaginary part $ \sigma _{2}$ ($\omega$) of the THz conductivity at various temperatures. A black dashed line highlights the curve at 40 K, and a bold red line highlights the curve at 2 K. The inset in (a) shows the temperature-dependent $ \sigma _{1}$ at the fixed frequency 1.0 THz, with the peak marked by a dashed line. (c, d) Normalized (c) real part $\sigma_1/\sigma_{1\mathrm{n}}$ and imaginary part $\sigma_2/\sigma_{2\mathrm{n}}$ of the conductivity as functions of temperature at selected frequencies, where $\sigma_{1\mathrm{n}}$ and $\sigma_{2\mathrm{n}}$ are the normal-state value at 40 K. The arrow in (c) marks the trend of the broad coherence peak. Panel (d) includes a gray dashed line depicting the result of the Mattis-Bardeen simulation at 0.3 THz. The corresponding equations are provided in the Supplementary Materials.} 
	\label{Figure2}
\end{figure}

Figures \ref{Figure2}a,b show the complex optical conductivity $\sigma(\omega)=\sigma_1(\omega)+i\sigma_2(\omega)$ extracted from the time‑domain data via model‑independent Fresnel analysis (Methods). At $T \gg T_\mathrm{c}^{\mathrm{onset}}$, $\sigma_1(\omega)$ exhibits a dip near 1 THz and a hump around 1.8 THz, while $\sigma_2(\omega)$ crosses zero at high frequency, deviating from a simple Drude response. We attribute this dip–hump structure to diffraction of the THz beam through the restricted effective aperture ($\simeq$ 1.6 mm, estimated from the distance between two opposing electrodes in the inset of  Fig. \ref{Figure1}c), as the beam waist ($\leq$ 2.6 mm, measured in Fig. \ref{THzspot}) exceeds this aperture. Consistent with this interpretation, the feature vanishes in batch \#3, where the effective aperture is increased to 3 mm with reduced electrode coverage (Fig. \ref{S3_sigma1}d), and a Drude-like response is recovered (Fig. \ref{S3_sigma1}c). However, in batch \#3, our THz spectroscopy measurements show a negligible superconductivity signature, despite identical sample growth conditions and resistivity measurements indicating similar $T_c$ values (comparisons for batches \#1–3 are provided in the supplementary materials and Fig. \ref{S3_sigma1}). Therefore, we focus on batch \#1, where the electrode geometry allows van der Pauw resistivity measurements in the central region of the film (inset of Fig. \ref{Figure1}c), and the THz spectra display clear superconductivity-related spectral-weight redistribution. Upon cooling toward $T_\mathrm{c}$, $\sigma_1(\omega)$ increases broadly while largely preserving its frequency dependence. Entering the superconducting state, part of spectral weight in $\sigma_1(\omega)$ is missed and $\sigma_2(\omega)$ develops an approximately $1/\omega$ divergence, unambiguously signaling condensate formation and transfer of spectral weight to the zero-frequency delta function. However, $\sigma_1(\omega)$ exhibits anomalous behavior that cannot be readily explained by either $s$- or $d$-wave superconductivity. The broad suppression from 0.5 THz to 2 THz below $T_\mathrm{c}^{\mathrm{onset}}$ resembles an $s$-wave response (Fig. \ref{Figure2}a), and suggests a gap scale above our measurement window. Moreover, $\sigma_1(\omega)$ retains a magnitude comparable to the normal state even at 2 K (Fig. \ref{Figure2}a), implying that approximately 65\% of carriers (estimated by the ratio of spectral weight $\frac{\int_{0.5}^{2} \sigma_{\text{1,2 K}}(\omega) \, d\omega}{\int_{0.5}^{2} \sigma_{\text{1,40 K}}(\omega) \, d\omega}$) remain uncondensed—reminiscent of residual absorption observed in cuprates \cite{Bilbro2011}. The imaginary part of THz conductivity $\sigma_2(\omega)$ (Fig. \ref{Figure2}b) is directly linked to the London penetration depth through the equation $c^2/\lambda_L^2 =\omega_{ps}^2=4\pi\omega\sigma_2(\omega)$, where $\omega_{ps}$ represents the plasma frequency of the superconducting condensate. Using the low-frequency limit, we estimate $\lambda_L\simeq 620$ nm at 2 K. This value is smaller than that inferred from kHz-range mutual-inductance measurements on (La,Pr)$_3$Ni$_2$O$_7$ (4 $\mu$m) with lower $T_\text{c}$ \cite{Zhou2025} but is much larger than those of elemental superconductors ($\simeq$ 50 nm) \cite{PhysRevB.50.6307,Kozhevnikov2013}, YBa$_2$Cu$_3$O$_7$ ($\simeq$ 200 nm) \cite{Mller1990,Pimenov1999}, and Ba$_{0.6}$K$_{0.4}$Fe$_{2}$As$_2$ ($\simeq$ 200 nm) \cite{Liprl2008}. The corresponding superfluid density is therefore almost an order of magnitude smaller than in cuprate and iron-based superconductors. As discussed below, the large residual quasiparticle response and reduced superfluid density impose strong constraints on the pairing symmetry.

To constrain the superconducting order-parameter symmetry, we analyzed the temperature dependence of the low-frequency THz conductivity and searched for a coherence peak, a hallmark of $s$-wave superconductors that is absent in $d$-wave cuprates \cite{PhysRevB.50.6307,Bonn1992}. Fig \ref{Figure2}c,d show $\sigma_1(T)$ and $\sigma_2(T)$ at selected frequencies. At $\omega=0.3$ THz, $\sigma_1(T)$ increases just below $T_\mathrm{c}$; with increasing frequency, the peak broadens while its position remains approximately unchanged. Near $T_\text{c}$, both coherence factors and superconducting fluctuations contribute to the enhancement of $\sigma_1(\omega,T)$. Simulations (Fig. \ref{MB_sim}) reproduce this feature, and the measured $\sigma_1(T)$ enhancement closely follows the simulated curve near $T_\mathrm{c}^{\mathrm{onset}}$, supporting its identification as a coherence peak. The measured peak is, however, much weaker than in the simulations (Fig. \ref{MB_sim}), consistent with substantial disorder. Although the coherence peak indicates an $s$-wave-like gap in the regions of the Brillouin zone probed by THz photons, the sizeable residual conductivity $\sigma_1(T \to 0)/\sigma_1(T_\mathrm{c}^{\mathrm{onset}}) \simeq 0.65$ in Fig. \ref{Figure2}a,c is incompatible with a conventional, fully gapped $s$-wave scenario, where $\sigma_1(0)/\sigma_1(T_\mathrm{c}^{\mathrm{onset}})$ is expected to be zero(see the simulation for $s$-wave superconductivity in Fig. \ref{MB_sim}b). In addition, $\sigma_2(T\rightarrow 0)/\sigma_2(T_\mathrm{c}^{\mathrm{onset}})\simeq 5$ at 0.3 THz is only about 25\% of the simulated $s$-wave value (Fig. \ref{Figure2}d). This substantial residual absorption as $T \to 0$ and the anomalously low superfluid density inferred from the large penetration depth ($\lambda_L \simeq $ 620 nm) raise the question of why so many carriers fail to condense. A plausible origin is pair-breaking scattering induced by intrinsic disorder and oxygen defects, which are inevitable in concurrent films. While such disorder typically has a minor effect on the low-frequency conductivity of conventional $s$-wave superconductors, it can severely deplete the superfluid density in $s_\pm$ candidates, where interband impurity scattering mixes hole and electron states with opposite values of the order parameter, which should act as pair-breaking in the same way as a magnetic impurity in $s$-wave superconductor \cite{Mazin2008}, thereby generating the observed residual absorption \cite{Aguilar2010}. An $s_{\pm}$ scenario is further supported by the established multiband electronic structure of (La,Pr)$_3$Ni$_2$O$_7$ and is consistent with spin-fluctuation-mediated pairing, as suggested by a growing body of experimental and theoretical studies \cite{Zhang2023,Shi2025Spin,Meng2024,Zhang2024NC,Liu2023,Gu2025,Tian2024}.

\vspace{18pt}
\noindent {\large \textbf{Anomalous normal state THz nonlinearity up to 100 K}}

Having identified bulk $s_{\pm}$-like superconductivity in (La,Pr)$_3$Ni$_2$O$_7$ films from the linear THz response, we now investigate their nonlinear THz response. This technique is sensitive to superconducting transitions, above-$T_\mathrm{c}$ fluctuations, and the coupling between the Higgs mode and other collective excitations in superconductors \cite{Matsunaga2014,Matsunaga2013,Matsunaga2012,Kovalev2021,Wang2022,Yuan2024,Chu2023,Kim2024}. Using this response, we further substantiate the superconducting state and probe its interplay with potential non-trivial normal states, such as the strange‑metal behaviour inferred from resistivity measurements \cite{Zhang2024high} or a pseudogap phase suggested by ARPES measurements \cite{shen2025LPSNO}.

\begin{figure}
	\includegraphics[width=1\columnwidth]{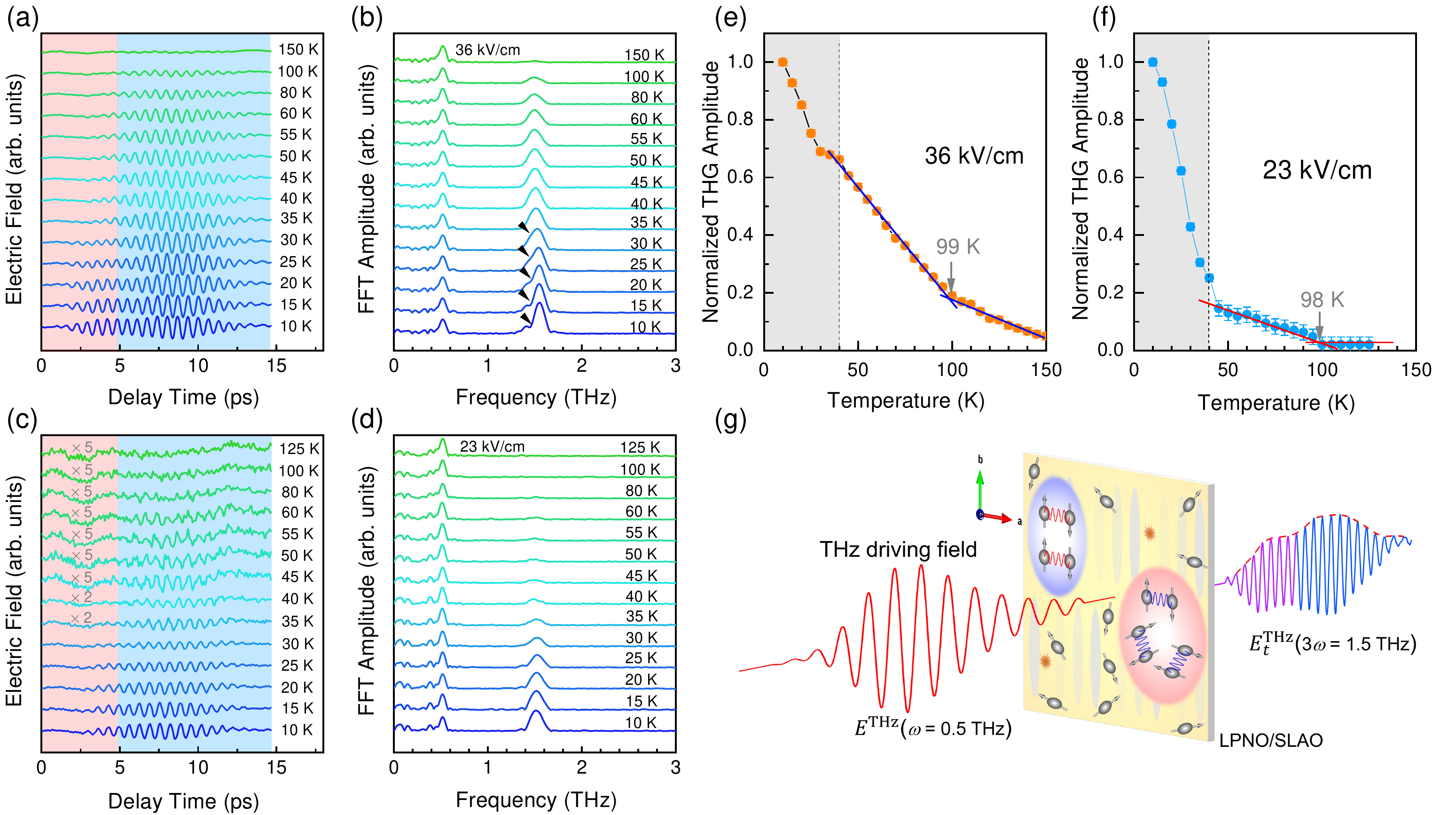}
	\caption{\textbf{Signatures of the anomalous normal state and superconductivity observed in the THG response at 0.5 THz.} (a, c) The real-time waveforms of the THG signal after a narrow band-pass filter (center frequency: 1.5 THz, i.e., $3\omega$) at selected temperatures, under driving field strengths of (a) 36 kV/cm, (c) 23 kV/cm. The fundamental pulse at 0.5 THz was suppressed by a band-stop filter (0.4–0.6 THz). The shaded orange and blue areas highlight the emergent THG waveform below 40 K and the primary THG component persisting up to 100 K, respectively. The signal above 30 K in panel (c) was amplified for clarity. (b, d) Corresponding fast Fourier transform (FFT) spectra of the transmitted pulses shown in (a) and (c). The kink in the THG spectrum is indicated by black arrows. (e, f) Temperature dependence of the THG amplitude under different driving field strengths. The amplitude increases rapidly below the superconducting transition at $T_\text{c}^{\mathrm{onset}} \simeq 40$ K (gray shaded region) and exhibits a distinct kink near 100 K (marked by gray arrows), which is delineated by two linear regimes (straight solid lines). (g) Schematic illustration of the proposed microscopic contributions to the THG signal. The blue region represents condensed Cooper pairs. Uncondensed electron pairs, potentially bound by a non-BCS mechanism, are depicted by the red region, while single black spheres denote unpaired electrons. Defect sites and an electronic stripe phase are indicated by orange stars and blue stripes, respectively. All these components may collectively contribute to the observed nonlinear response.
	}
	\label{Figure3}
\end{figure}

We first performed comparative measurements at 10 K on an LPNO/SLAO sample and a bare SLAO substrate with identical electrodes (Fig. \ref{Exclu_sub}). The THG signal appeared exclusively in the LPNO/SLAO sample, confirming that the response originates solely from the film. We then measured the temperature dependence of the THG from the (La,Pr)$_3$Ni$_2$O$_7$ film driven by a 0.5-THz multi-cycle pulse. At high peak field ($E_{\text{peak}} = 36$ kV cm$^{-1}$), the transmitted waveforms revealed two distinct wavelets at low temperatures (Fig. \ref{Figure3}a; orange and blue shading). As the temperature increased, the wavelet in the early time window wakens and becomes invisible above 40 K, while the second wavelet persists in the normal state, up to temperature higher than 100 K. Corresponding to these wavelets, the fast Fourier transformed THG spectra (Fig. \ref{Figure3}b) at 10 K displayed a kink (marked by a black arrow) indicative of a beat frequency, which vanished above 40 K, linking this feature to the superconducting transition. Concomitantly, the field-scaling exponent $n$ of the THG amplitude (defined by $A_{\text{THG}} \propto E_{\text{peak}}^n$) dropped abruptly below 40 K and deviated from the cubic scaling expected in the normal state (Fig. \ref{Flu_law}). Reducing the drive field to 23 kV cm$^{-1}$ suppresses the two-wavelets structure (Fig. \ref{Figure3}c), leaving only a weak low-temperature asymmetry in the frequency domain (Fig. \ref{Figure3}d). The field-induced spectral splitting resembles observations in optimally doped YBa$_2$Cu$_3$O$_{6+x}$ \cite{Yuan2024}. Furthermore, reducing the drive field suppressed THG above $T_\mathrm{c}^{\mathrm{onset}}$, while comparatively enhancing the low-temperature response. These observations clearly indicate two distinct contributions to the THG signal below $T_\mathrm{c}^{\mathrm{onset}}$: one from superconductivity and the other from normal-state nonlinearity invaded into the superconducting state. 

To explicitly track the temperature evolution of the two contributions, we plotted the normalized THG amplitude, $A_{\text{norm}}(T)$, at several field strengths (Fig. \ref{Figure3}e, f). At both high and low fields, $A_{\mathrm{norm}}(T)$ rises sharply below the transport onset temperature $T_\mathrm{c}^{\mathrm{onset}}\simeq 40$ K (grey shading). The temperature dependence nevertheless shows two anomalies. First, the expected THG intensity enhancement arising from the $2\Delta(T) = 2\omega$ resonance below $T_\text{c}$ \cite{Matsunaga2014,Wang2022,Kovalev2021,Yuan2024} is absent in (La,Pr)$_3$Ni$_2$O$_7$, likely due to intrinsic disorder, consistent with the large uncondensed electron population revealed by THz linear spectroscopy. Second, above $T_\text{c}^{\text{onset}}$, the signal persists, similar to the above-$T_\text{c}$ THG signal observed in cuprates and disordered NbN \cite{Katsumi2018,Chu2020,Yuan2024,wang2025anomal}, but in contrast to clean $s$-wave NbN \cite{Matsunaga2013,Matsunaga2014,Wang2022} and $s_\pm$ Co‑doped BaFe$_2$As$_2$ \cite{JGWang_iron_based_Higgs}, where the nonlinear response vanishes promptly at $T_\mathrm{c}$. In addition, $A_{\text{norm}}(T)$ shows a reproducible kink near 99 K (98 K) for $E_{\text{peak}}=36$ (25) kV cm$^{-1}$. This feature is confirmed by measurements at a second drive frequency of 0.42 THz (Fig. \ref{S1_0.42}), where the same characteristic temperature is observed despite different field dependencies in the superconducting and normal states. Notably, recent ARPES studies reveal a pseudogap emerging below 100 K \cite{shen2025LPSNO} in similarly grown (La,Pr,Sm)$_3$Ni$_2$O$_7$ films, pointing to a common normal‑state energy scale identified by independent probes. These results raise the question of which degrees of freedom—or incipient order—give rise to the anomalous nonlinear response in the normal state.

Further insights come from sample-dependent measurements. Although the temperature‑ dependent resistivity of three films (batches \#1–\#3) is similar, their THz responses differ substantially. Batch \#1 shows strong THG at low temperature, whereas batch \#2 exhibits a much weaker THG amplitude at 5 K (Fig. \ref{S2_0.42}) and a reduced superfluid density in linear THz spectroscopy. Notably, the THG signal in batch \#2 is negligible above $T_\mathrm{c}^{\mathrm{onset}}$, and batch \#3 shows neither a detectable superconducting response nor any THG signal over the accessible temperature range. These trends demonstrate that the normal‑state nonlinear response correlates with the robustness of superconductivity at low temperature.

Given the pronounced electronic inhomogeneity of the (La,Pr)$_3$Ni$_2$O$_7$ films, we propose a granular superconducting scenario in which superconducting regions are embedded within a metallic normal‑state background (Fig. \ref{Figure3}g). This picture naturally accounts for both the large uncondensed carrier population revealed by linear THz spectroscopy and the beating feature observed in the nonlinear response. The character of the normal state, however, remains unresolved. One possibility is a trivial metallic state with strong scattering, exhibiting a nonlinear response similar to that of disordered NbN \cite{wang2025anomal}. In this limit, the films would be far from optimal superconductivity and potentially close to a superconductor–insulator transition. An alternative possibility is that the normal state hosts a pseudogap phase. Our sample‑dependent measurements show that the strength of the normal‑state nonlinearity correlates closely with the robustness of low‑temperature superconductivity. In parallel, recent transport studies suggest that enhanced superconductivity coincides with non‑Fermi‑liquid, strange‑metal behaviour in the normal state \cite{zhou2025onset60k}. Together, these observations raise the possibility that a strange‑metal regime develops at elevated temperature, within which a pseudogap sets in below a crossover temperature $T_{\mathrm{THG}}^{\mathrm{onset}}\sim100$ K. This temperature is consistently identified by surface‑sensitive ARPES \cite{shen2025LPSNO} and by the bulk nonlinear THz response. In such a scenario, the anomalous normal‑state nonlinearity would originate from a distinct order. Distinguishing between these possibilities will require systematic studies on cleaner, more homogeneous films.

\vspace{18pt}
\noindent {\large \textbf{Discussion and outlook}}

Our results identify several key aspects of pairing in superconducting RP nickelates. First, the coexistence of a low‑frequency coherence peak with substantial residual conductivity is consistent with sign‑changing $s_{\pm}$ pairing. Second, the large London penetration depth and the absence of a Higgs‑mode resonance point to strong disorder and electronic inhomogeneity. Third, the anomalous normal-state THz nonlinearity indicates the presence of an additional phase-potentially a pseudogap, a strange metal, or a combination of both—persisting above $T_\mathrm{c}$. These features distinguish RP nickelates from the better‑established cuprate and iron‑based superconductors. In bilayer nickelates, the data support a sign‑reversing $s_{\pm}$‑like pairing state, in contrast to the nodal $d_{x^2-y^2}$ pairing of cuprates and reminiscent of multiband $s_{\pm}$ pairing in iron‑based systems. Proposed microscopic scenarios further suggest that localized $d_{z^2}$ electrons may form pre‑paired states via interlayer antiferromagnetic superexchange, which subsequently hybridize with itinerant $d_{x^2-y^2}$ carriers to establish phase-coherent superconductivity. This mechanism differs from nearest-neighbour Heisenberg exchange in cuprates and from predominantly interband spin‑fluctuation pairing in many iron-based materials.

The nature of the normal state preceding superconductivity in RP nickelates remains less well understood. Our nonlinear THz measurements point to an anomalous phase above $T_\mathrm{c}$ that is likely a pseudogap, akin to that commonly intertwined with superconductivity in cuprates. By contrast, in iron‑based superconductors superconductivity typically emerges as long‑range antiferromagnetic order is suppressed and is not necessarily accompanied by a pseudogap. By revealing a candidate exotic normal state together with strong disorder effects, our work narrows the range of viable theoretical descriptions. Nonetheless, the microscopic origin of disorder and inhomogeneity in the present films remains unclear, underscoring the need for improved growth control. Future experiments on cleaner samples, combined with complementary probes—most notably ARPES performed on the same films—will be crucial for tracking the evolution of superfluid density, gap structure and pseudogap behaviour with strain and doping, thereby further constraining the pairing mechanism.

\vspace{18pt}
\noindent {\large \textbf{Methods}}

\textbf{Sample growth.} High-quality (La,Pr)$_3$Ni$_2$O$_{7}$ films with a (La,Pr)$_2$NiO$_{4}$ buffer layer were grown on the as-received SrLaAlO$ _{4} $ (001) substrates (MTI, China) by the Gigantic-Oxidative Atomically Layer-by-Layer Epitaxy (GAE) method \cite{10.1093/nsr/nwae429}. Notably, the film did not undergo any post-annealing treatment.

\begin{sloppypar}
\textbf{The linear terahertz spectroscopy measurements.} Terahertz conductivity measurements were performed using a pair of photoconductive antennas. The temporal electric-field waveform was recorded by fast scan technology. The complex transmission function was obtained by comparing the Fourier‑transformed signals transmitted through the (La,Pr)$_3$Ni$_2$O$_7$/SrLaAlO$_4$ sample and a bare SrLaAlO$_4$ reference. The complex refractive index $\tilde{n}$ of the film was extracted by solving the standard thin‑film transmission expression  $\tilde{T}(\omega) = \frac{2\tilde{n}(1+\tilde{n}_{\text{sub}}) \exp\left[i\frac{(\tilde{n}-1)\omega d}{c}\right] \exp\left[i\frac{(\tilde{n}_{\text{sub}}-1)\omega \Delta L}{c}\right]}{(1+\tilde{n})(\tilde{n}+\tilde{n}_{\text{sub}})-(\tilde{n}-1)(\tilde{n}-\tilde{n}_{\text{sub}})\exp\left(i\frac{2\tilde{n}\omega d}{c}\right)} $ where $\tilde{n}_{sub}$ is the complex refractive index of substrate, $d$ is the thin film thickness and $\Delta\it{L}$ is the little thickness difference between SrLaAlO$ _{4} $ substrate of sample and the reference. The precise value of $\Delta\it{L}$ was determined by finding the minimum of $\left|\frac{\tilde{t}_{II}}{\tilde{t}_{I}}(\tilde{n}_{sub}-1)-2\tilde{n}_{sub}\tilde{t}_{I}exp[ik\Delta L(1+\tilde{n}_{sub})]+(1+\tilde{n}_{sub})exp(ik\Delta L2\tilde{n}_{sub}) \right|$, where $\tilde{t}_{I}=\tilde{E}_{\mathrm{sam}}/\tilde{E}_{\mathrm{ref}}$ and $\tilde{t}_{II}$ denotes the first echo of $\tilde{t}_{I}$ \cite{Krewer2018}. The complex optical conductivity was then obtained directly from $\tilde{n}$ via $\tilde{n}^{2}=1+i\tilde{\sigma}(\omega)/(\omega\varepsilon_{0})$, without invoking a Kramers–Kronig transformation. To minimize oxygen loss, the sample was rapidly cooled to low temperature in an Helium-filled chamber.
\end{sloppypar}

\textbf{THG measurements.} Strong-field broadband terahertz pulses were generated using a tilted-pulse-front scheme in a LiNbO$ _{3} $ crystal, driven by an amplified Ti:Sapphire laser system (central wavelength: 800 nm, pulse duration: 35 fs, repetition rate: 1 kHz). Narrowband, multicycle terahertz pulses were obtained by passing the broadband output through a set of band-pass filters (centered at 0.5 THz) placed before the sample. To weaken the fundamental pulse intensity and enhance the detection sensitivity for the THG signal, an additional band-pass filter (centered at 1.5 THz) was inserted after the sample. The waveforms of the terahertz pulses generated by sample are measured by electro-optic sampling with a 2-mm-thick ZnTe crystal. To mitigate oxygen loss from the sample during thermal cycling ($>$ 200 K) in vacuum, a gas closed-cycle cryostat is employed. The sample was first cooled rapidly to 120 K in an oxygen-filled chamber, then the chamber was purged with helium before further cooling to the base temperature of 5 K. The warming cycle followed the reverse sequence.

\textbf{Transport measurements.} Temperature-dependent resistivity was measured in a closed-cycle helium-free cryostat with the lowest temperature of approximately 1.8 K and a magnetic field of up to 14 T. Platinum Hall bar electrodes were deposited onto 5$\times$5 mm$ ^{2} $ samples via magnetron sputtering through a pre-patterned hard-shadow mask. To minimize oxygen loss, the sample was rapidly cooled below 200 K within 10 minutes, to reduce exposure time under conditions conducive to oxygen desorption.

\medbreak
\textbf{Data availability} Source data are provided with this paper. Data generated or analyzed during this study are included in the Article and the Supplementary Information.

\vspace{18pt}

\bibliographystyle{apsrev4-2}

\bibliography{327filmrefs}

\medbreak
\vspace{18pt}
\noindent {\textbf{Acknowledgments}} This work was supported by the National Natural Science Foundation of China (Grants No. 12488201, 92265112, 12374455, 12574349, 52388201, and 92565303), the National Key Research and Development Program of China (Grants No. 2024YFA1408700, 2022YFA1403901, 2021YFA1400201, 2022YFA1403101 and 2024YFA1408101), the Guangdong Major Project of Basic Research (Grant No.~2025B0303000004), the Guangdong Provincial Quantum Science Strategic Initiative (Grants No.~GDZX2501001, GDZX2401004, and GDZX2201001), the Shenzhen Science and Technology Program (Grant No.~KQTD20240729102026004), the Shenzhen Municipal Funding Co-Construction Program Project (Grants No.~SZZX2301004 and SZZX2401001), the National Natural Science Foundation of China (Grant No.~12504166), the Synergetic Extreme Condition User Facility (SECUF, https://cstr.cn/31123.02.SECUF), and the CAS Superconducting Research Project under Grant No.[SCZX-0101].

\medbreak
\noindent {\textbf{Author contributions}} Qi-Kun Xue, ZhuoyuChen, Tao Dong and Nanlin Wang conceived the project. Shuxiang Xu and Xinbo Wang performed the linear terahertz spectroscopy and THG measurements with help from Hao Wang, Tianyi Wu, and Liyu Shi at the Synergetic Extreme Condition User Facility. Guangdi Zhou prepared and characterized the (La,Pr)$_3$Ni$_2$O$_{7}$ films with the help of Wei Wang and Haoliang Huang under the supervise of Zhuoyu Chen and Jinfeng Jia. Hao Wang calculated the linear optical conductivity. Shuxiang Xu, Tao Dong and Nanlin Wang analyzed the data and wrote the manuscript with inputs from all authors.
\medbreak
\noindent {\textbf{Competing interests}} The authors declare no competing financial interests. 
\medbreak
\noindent {\textbf{Additional information}} Supplementary information available. 
\clearpage

\clearpage
\setcounter{figure}{0}
\setcounter{equation}{0}
\renewcommand{\thefigure}{S\arabic{figure}}
\begin{center}
	\textbf{\Large Supplemental material for "Multimodal Terahertz Spectroscopy of the Pairing Symmetry and Normal-State Pseudogap in (La,Pr)$_3$Ni$_2$O$_7$ Films"}
\end{center}

\section{Comprehensive comparison of terahertz conductivity and resistivity across different sample batches}
\begin{figure}[htbp]
	\centering
	\includegraphics[width=0.75\linewidth]{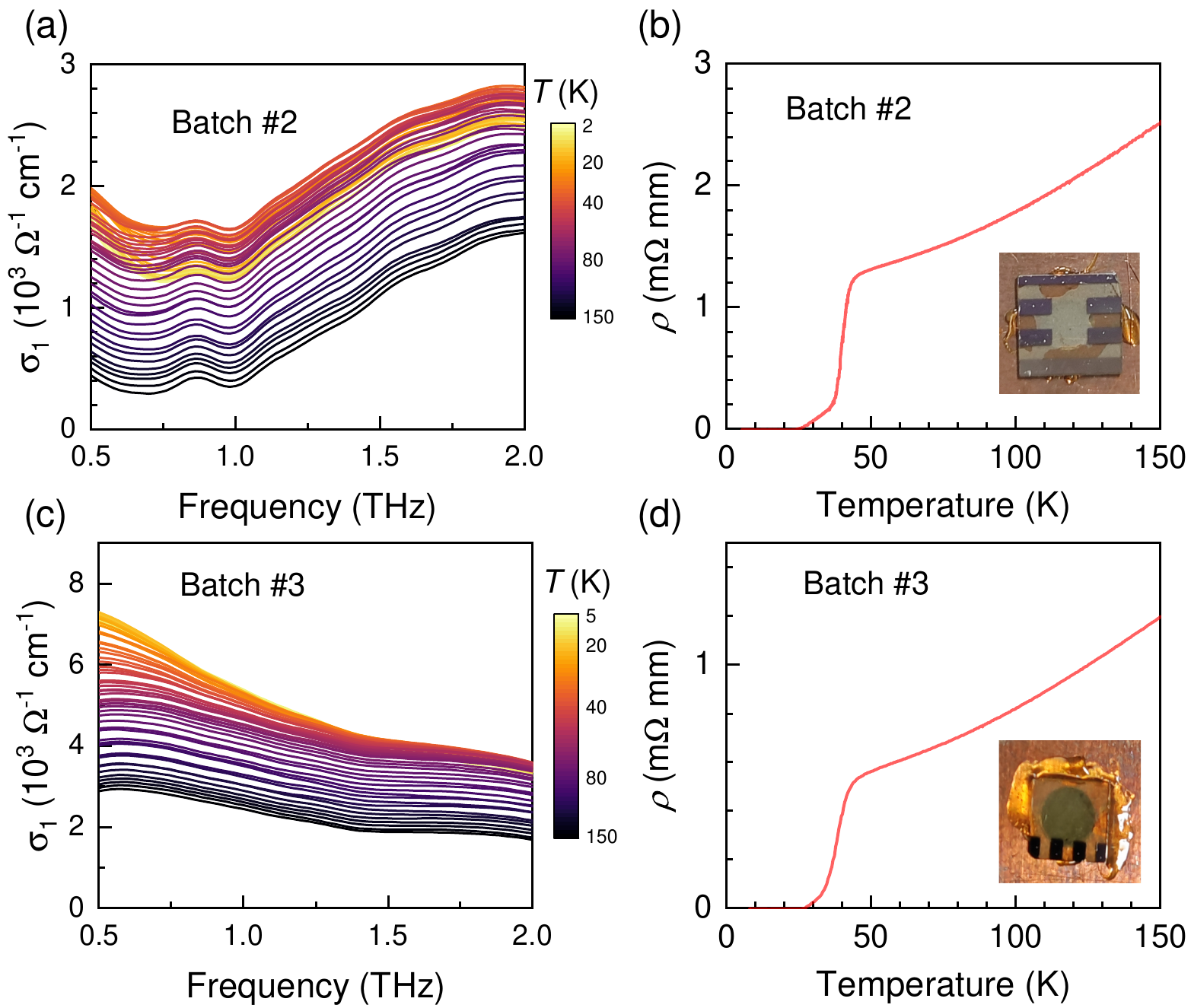}
	\caption{(a) Frequency-dependent real part $ \sigma _{1}$ ($\omega$) of THz conductivity for batch~\#2 (La,Pr)$_3$Ni$_2$O$_{7}$/SrLaAlO$_4$ sample at various temperatures. (b) Temperature-dependent resistivity of the batch~\#2 sample. The inset shows the configuration of the six broad Pt electrodes. (c) $\sigma_1(\omega)$ for batch \#3 sample at various temperatures. (d) Temperature-dependent resistivity of the batch~\#3 sample. The inset shows the configuration of the parallel Pt electrodes. The color variation of the samples is due to differences in the photography angle and light intensity.}
	\label{S3_sigma1}
\end{figure}

To verify the reproducibility of the terahertz spectroscopy in the superconducting state, we performed THz measurements on a second sample (batch~\#2) with the same electrode geometry as batch~\#1. The real part of the THz conductivity is presented in Fig.~\ref{S3_sigma1}(a). Below the onset transition temperature $T_\text{c}^{\text{onset}}$ (42~K), a reduction in spectral weight is observed, although it is less pronounced than in batch~\#1. This difference may be attributed to a broader superconducting transition tail in batch~\#2, as seen in Fig.~\ref{S3_sigma1}(b). Furthermore, the THz conductivity spectrum of batch~\#2 exhibits a similar dip-like feature around 1~THz to that of the first sample. In both batches, the relatively large electrode pads limit the effective optical aperture to approximately 1.6~mm, which can readily introduce diffraction effects. To mitigate this, we fabricated a third sample (batch~\#3) with four narrow, parallel electrodes, increasing the clear aperture to about 3~mm. As anticipated, the dip feature vanishes completely, as demonstrated in Fig.~\ref{S3_sigma1}(c). However, despite clear signatures of superconductivity in the resistivity measurements (Fig.~\ref{S3_sigma1}(d)), no detectable loss of spectral weight is observed across the temperature range investigated. This discrepancy likely stems from the fact that the resistivity and THz conductivity measurements probe different regions of the sample. These results suggest that the current samples exhibit considerable spatial inhomogeneity on a macroscopic scale.

\section{The THz conductivity simulated by Mattis-Bardeen equation}
\begin{figure}[htbp]
	\centering
	\includegraphics[width=0.95\linewidth]{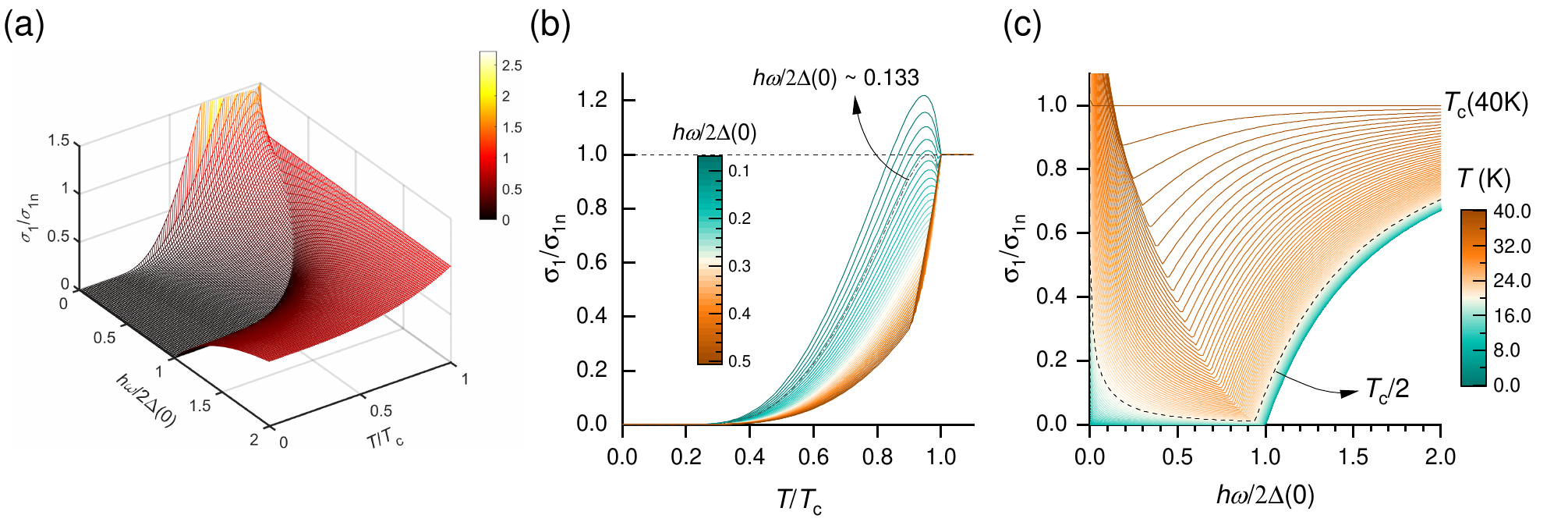}
	\caption{Simulated temperature and frequency dependence of the normalized real conductivity $\sigma_1/\sigma_n$ based on the Mattis-Bardeen theory. (a) Three-dimensional surface plot. (b) Temperature dependence at selected frequencies. (c) Frequency dependence at selected temperatures.}
	\label{MB_sim}
\end{figure}

For conventional $s$-wave superconductors, the real part of optical conductivity in the superconducting state can be described by the Mattis-Bardeen equation \cite{PhysRev.111.412}
\begin{equation}
	\frac{\sigma_{1s}}{\sigma_{1n}} = \frac{2}{\hbar\omega} \int_{\Delta}^{\infty} [f(E) - f(E + \hbar\omega)]g_1(E, E + \hbar\omega) dE + \frac{1}{\hbar\omega} \int_{\Delta - \hbar\omega}^{-\Delta} [1 - 2f(E + \hbar\omega)]g_1(E, E + \hbar\omega)dE,
\end{equation}
where $f(E) = 1/[1+\exp(E/T)]$ is the Fermi-Dirac distribution function, $E$ denotes the energy relative to the Fermi Level, and $\Delta$ is the superconducting energy gap. The coherence factor $g_1(E, E + \hbar\omega)$ is given by 
\begin{equation}
	g_1(E, E + \hbar\omega)= \frac{E^2+\Delta^2+\hbar\omega E}{u_1 u_2}, 
\end{equation}
with $u_1=(E^2-\Delta^2)^{1/2}$ and $u_2=[(E+\hbar\omega)^2-\Delta^2]^{1/2}$ being the corresponding Bogoliubov quasiparticle energies for states at $E$ and $E+\hbar\omega$, respectively. The THz conductivity was simulated using the Mattis-Bardeen formalism with $T_\text{c}$ = 40 K and $\Delta$ = 8 meV (Fig.~\ref{MB_sim}). This theory predicts that a coherence peak appears only at low frequencies, typically for $\hbar\omega / \Delta(0) < 0.1$ \cite{PhysRevB.50.6307}. In our simulation, the peak value falls below unity for $\hbar\omega / \Delta(0) < 0.133$, as shown in Fig.~\ref{MB_sim}(b).

\section{Characterization of the terahertz probe beam}
\begin{figure}[htbp]
	\centering
	\includegraphics[width=0.8\linewidth]{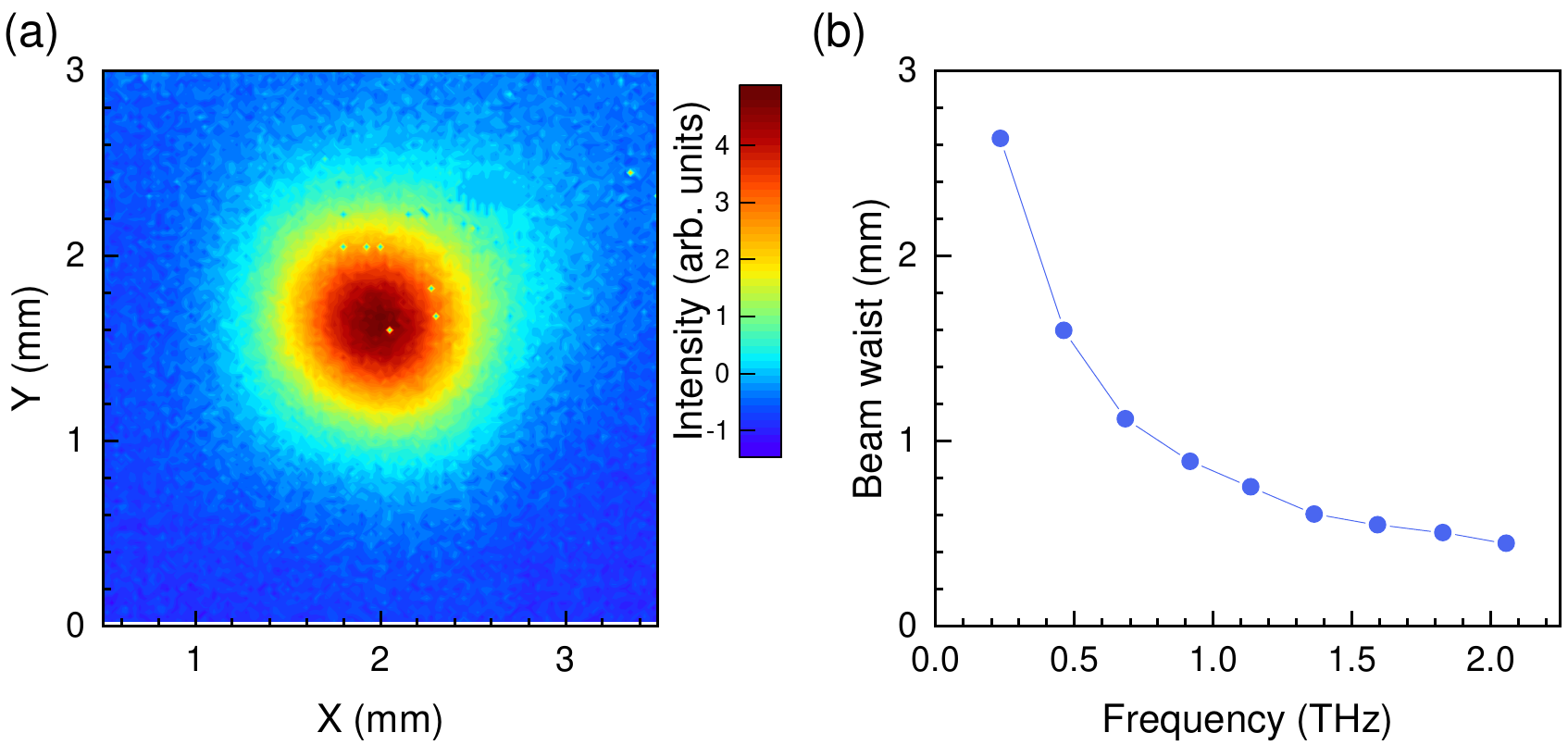}
	\caption{(a) Measured spatial intensity profile of the THz beam. (b) Beam waist diameter as a function of frequency. }
	\label{THzspot}
\end{figure}

The spatial intensity profile of the THz beam was characterized using a THz camera, as shown in Fig.~\ref{THzspot}(a). The beam diameter was determined by fitting the data to a Gaussian function, $I(r) = I_0 \exp\left(-\frac{2r^2}{w^2}\right)$, where $w$ denotes the beam waist radius. From this measurement, the $1/e^2$ beam diameter (\(2w\)) is approximately 1.65~mm. Additionally, the frequency dependence of the beam diameter was investigated using the knife-edge method. As shown in Fig.~\ref{THzspot}(b), the beam waist increases significantly as the frequency decreases below 1 THz.

\section{Comparing the nonlinear THz responses of the sample and the substrate.}
\begin{figure}[htbp]
	\centering
	\includegraphics[width=0.4\linewidth]{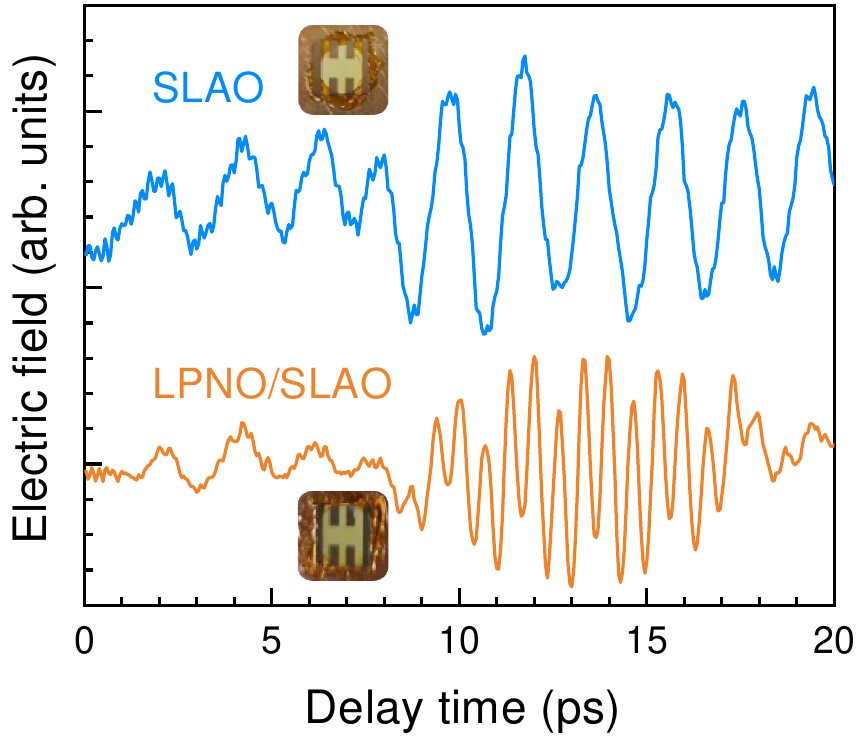}
	\caption{The nonlinear THz response for the (La,Pr)$_3$Ni$_2$O$_7$/SrLaAlO$_4$ sample (orange curve) and the bare SrLaAlO$_4$ substrate with electrodes (blue curve) under 0.5 THz excitation. The THG signal is observed exclusively in the sample.}
	\label{Exclu_sub}
    \end{figure}

In our experiments, electrical transport measurements were performed on the samples before spectroscopic characterization. To exclude any potential contribution from the Pt electrodes (30–100 nm thick) to the nonlinear optical response, we separately measured the third-order nonlinear terahertz response from the electrode-coated substrate and the sample. As shown in Fig.~\ref*{Exclu_sub}, under excitation by multicycle 0.5 THz pulses, a third-harmonic signal at 1.5 THz was detected only in the sample, with no signal observed from the substrate. This confirms that the electrodes do not contribute to the third-harmonic generation, and thus the measured nonlinearity originates entirely from the sample itself.
	
\section{Analysis of fluence-dependent THG intensity at low temperatures}
\begin{figure}[htbp]
		\centering
		\includegraphics[width=0.8\linewidth]{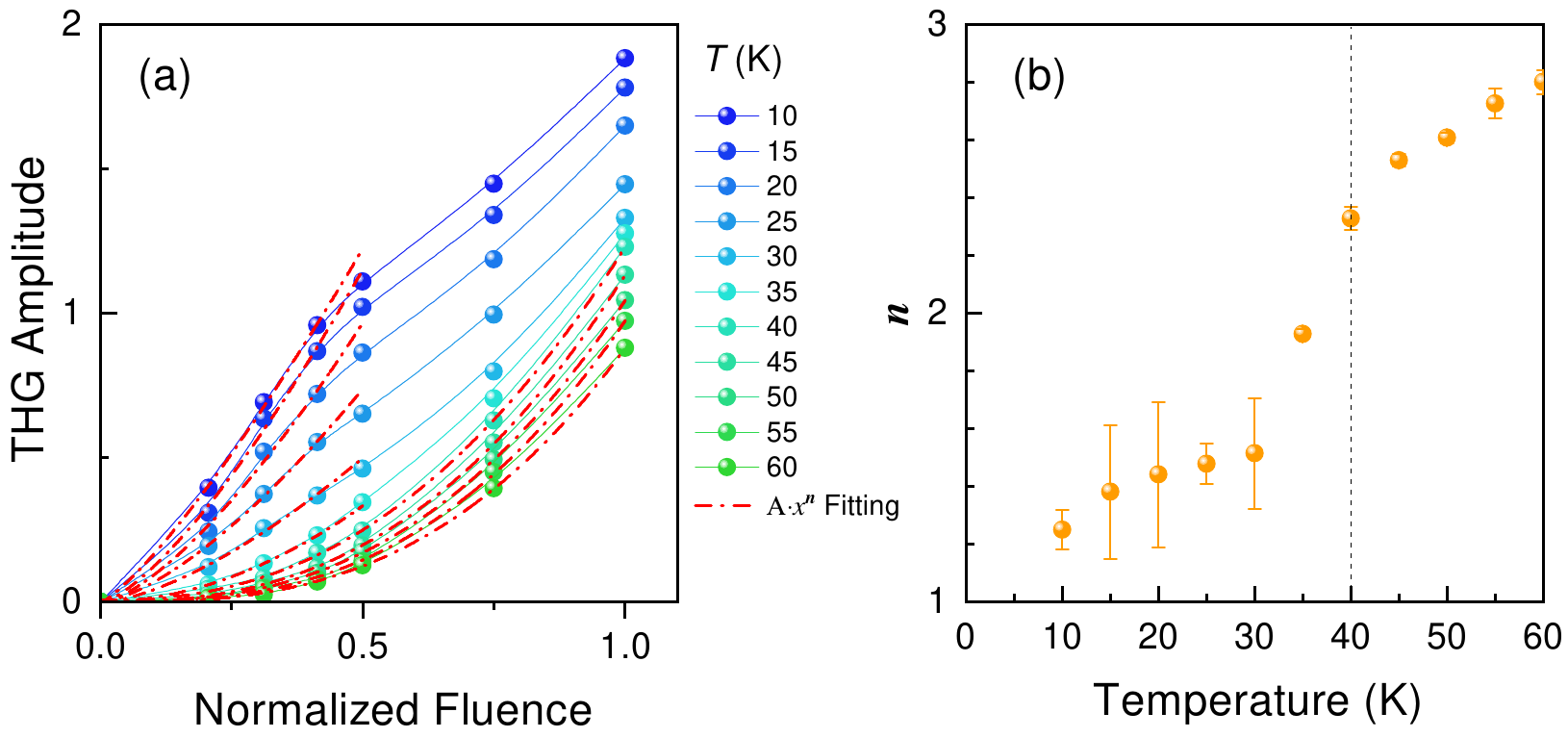}
		\caption{ (a) THG intensity $A_{\text{THG}}$ versus pump electric field strength $E_{\text{peak}}$ at various temperatures. The red dashed lines are fits to $A_{\text{THG}}=C \cdot E_{\text{peak}}^{n}$, where $C$ is a coefficient and $n$ is a power-law exponent. (b) Temperature dependence of the power-law exponent $n$. The black dashed line marks the $T_\text{c}^{\text{onset}} \simeq 40$ K.}
		\label{Flu_law}
	\end{figure}

The fluence dependence of the third-harmonic generation (THG) intensity was measured at low temperatures and fitted with a power-law function, as shown in Fig.~\ref{Flu_law}. The extracted exponent, $n$, increases gradually with temperature, exhibits a rapid rise near 40 K, and then slows, eventually approaching a value of 3. In conventional third-order nonlinear optics, a cubic dependence ($n=3$) is expected. The observed evolution of $n$ across the superconducting transition suggests a change in the dominant nonlinear mechanism, consistent with the emergence of the superconducting phase below 40 K.

\section{Reproducible ordered phases under 0.4 THz excitation}
\begin{figure}[htbp]
	\centering
	\includegraphics[width=0.8\linewidth]{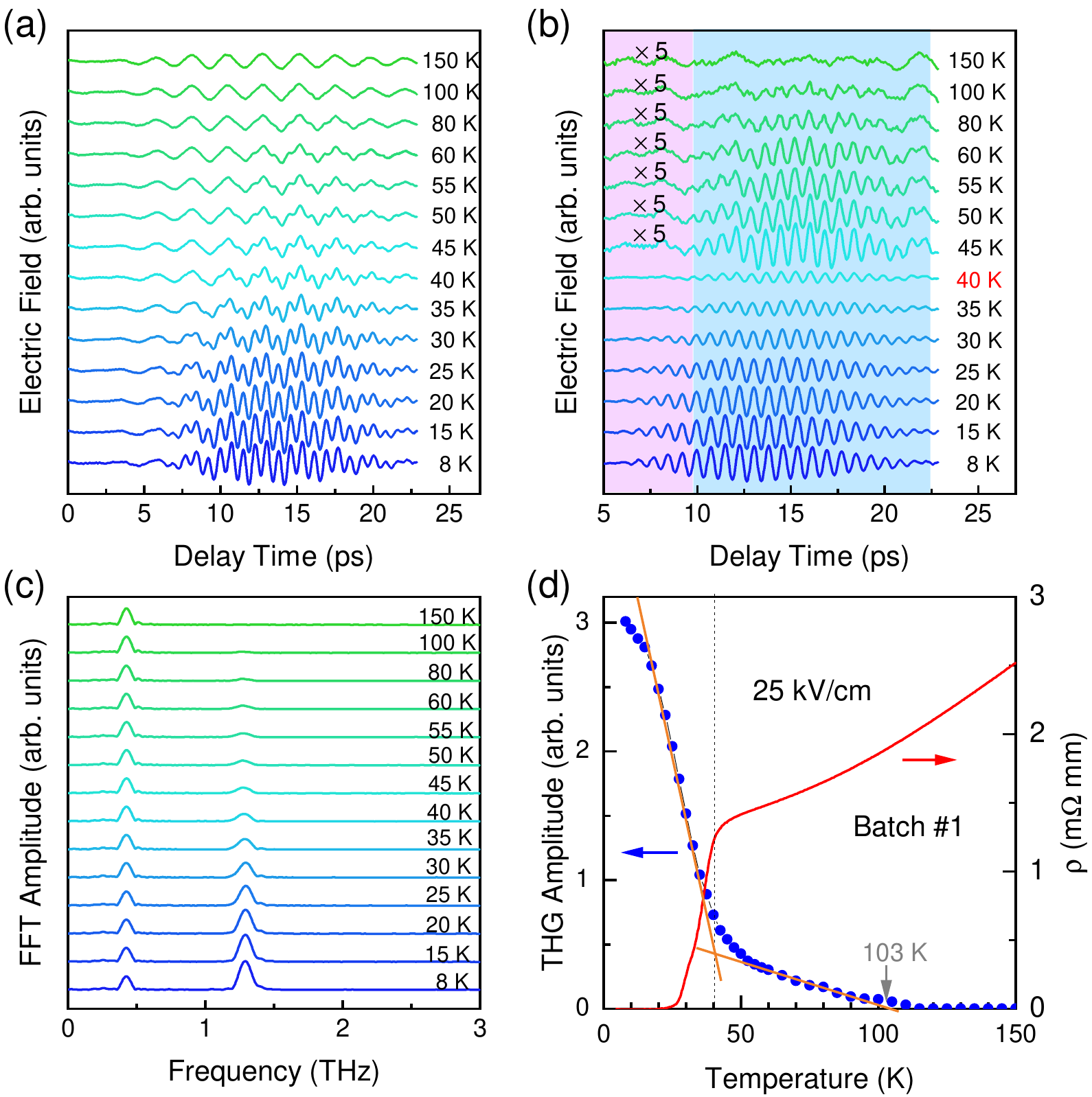}
	\caption{THG response of the batch \#1 LPNO/SLAO sample under 0.42 THz driving. (a) Temporal waveforms of the THG signal after a high frequency (3$ \omega $=1.26 THz) band-pass filter at selected temperatures. (b) THG signal derived from (a) following band-stop filtration (0.37-0.47 THz). The purple shaded region denotes the additional THG waveform emerging below 40 K, whereas the blue shaded area corresponds to the primary THG waveform, which persists up to 100 K. Signals above 40 K are amplified by a factor of five. (c) Fast Fourier transform amplitude spectra calculated from the waveforms in (a). (d) Blue symbols represent the THG amplitude obtained from Fourier analysis of the waveforms in (a). The red solid line indicates the temperature-dependent resistivity. The black dashed line marks the superconducting onset temperature $T_\text{c}^{\text{onset}} \simeq 40$ K, below which the THG amplitude exhibits a rapid enhancement. Orange lines delineate two distinct linear regimes, and the gray arrow identifies the temperature corresponding to the extrapolated intercept of the second linear segment. }
	\label{S1_0.42}
\end{figure}

Upon excitation with 0.42 THz radiation, a pronounced THG signal is also detected in the batch \#1 sample at low temperatures. Upon increasing the temperature, the THG signal gradually attenuates. It remains detectable above 40 K and eventually vanishes completely above 100 K, as depicted in Fig.~\ref{S1_0.42}(a). When the driving fundamental frequency is filtered out, an increase in the number of oscillations within the THG signal is observed clearly upon cooling below 40 K as shown in Fig.~\ref{S1_0.42}(b), suggestive of an emergent beat frequency. Fast Fourier Transform analysis of the data in Fig.~\ref{S1_0.42}(a) reveals the temperature-dependent spectra of the fundamental and the third-harmonic frequencies, shown in Fig.~\ref{S1_0.42}(c). At low temperatures, the THG intensity substantially exceeds that of the fundamental. Notably, while the fundamental signal persists across the measured temperature range, the THG signal diminishes rapidly with increasing temperature and eventually disappears completely above 100 K. By extracting the THG intensity and correlating it with the electrical transport curve, two distinct linear temperature-dependent regimes are resolved for the THG response. One intersection point coincides with the superconducting onset temperature 40 K and another occurs at approximately 103 K, consistent with the pseudogap onset temperature determined by ARPES measurements. These observations confirm that a third-harmonic nonlinear optical response persists under 0.42 THz excitation across the superconducting and pseudogap phases, furnishing additional evidence for the coexistence of two distinct ordered phases in the (La, Pr)$_3$Ni$_2$O$_7$ films.

\section{Weak THG response of complex orders in a low-quality sample}
\begin{figure}[htbp]
	\centering
	\includegraphics[width=0.8\linewidth]{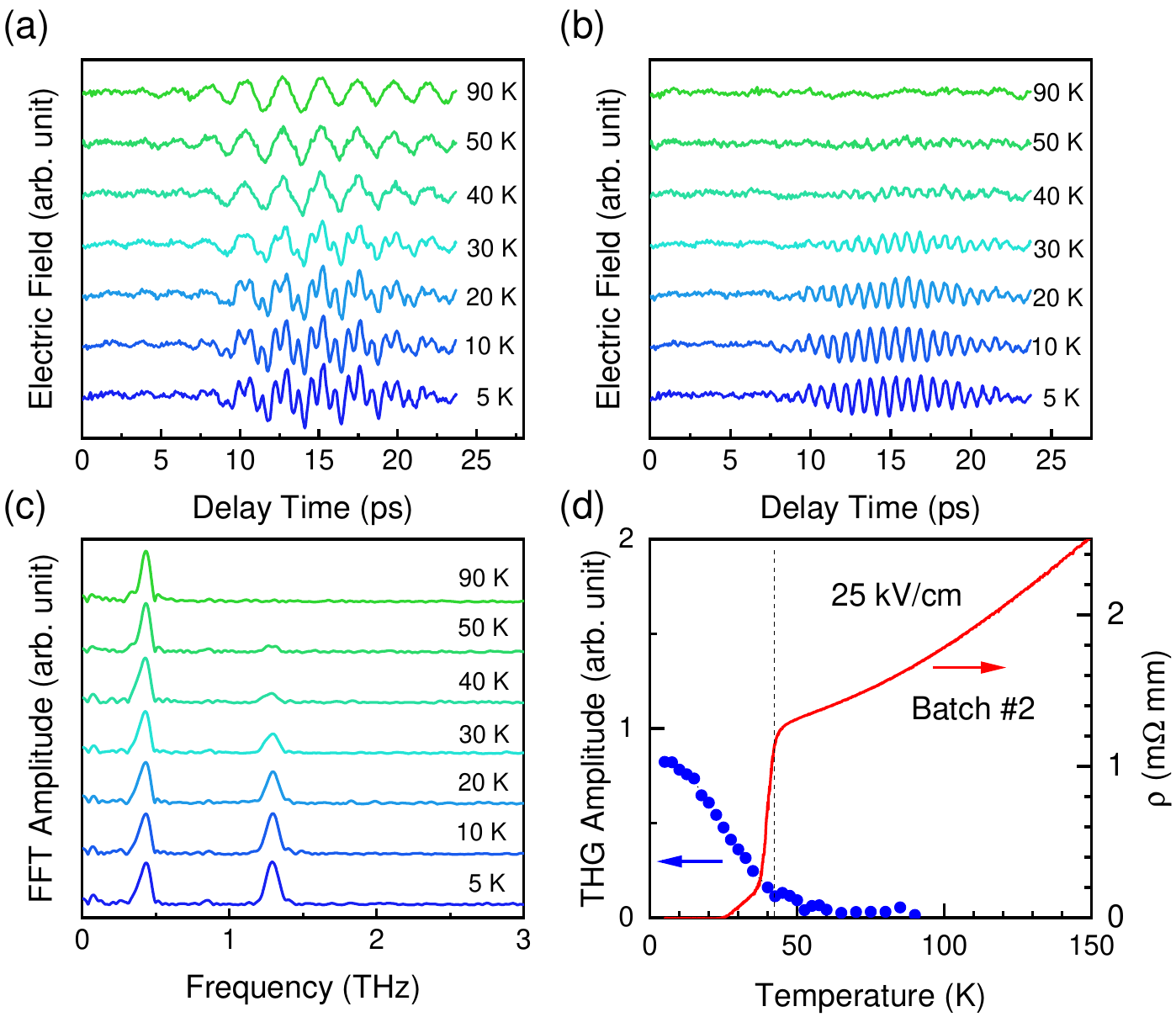}
	\caption{THG response of the batch \#2 LPNO/SLAO sample under 0.42~THz excitation. (a) The real-time waveforms of the THG signal after a high-frequency ($3\omega = 1.26$~THz) band-pass filter at selected temperatures. (b) THG signal obtained from (a) after additional band-stop filtering (0.37-0.47~THz). (c) Fast Fourier transform amplitude spectra derived from the waveforms in (a). (d) Blue dots denote the THG amplitude extracted from Fourier analysis of the waveforms in (a). The THG amplitude increases steeply below 40~K. The red solid line represents the temperature-dependent resistivity, and the black dashed line indicates the superconducting transition temperature $T_{\mathrm{c}}^{\text{onset}} \simeq 42$~K.}
	\label{S2_0.42}
\end{figure}

 The THG response in superconducting film samples exhibits a pronounced dependence on sample quality. To investigate this, we characterized an additional sample of inferior quality, which exhibited a superconducting onset temperature (42~K) comparable to that of the batch \#1 sample but with a broadened superconducting transition tail. Under 0.42~THz excitation, a THG signal was detectable at low temperatures; however, it attenuated rapidly upon warming and became undetectable above 50~K, as shown in Figs.~\ref{S2_0.42}(a-b). Spectral analysis via FFT further reveals that even at the lowest temperature 5 K, the THG intensity is nearly comparable to that of the fundamental frequency (Fig.~\ref{S2_0.42}(c)), indicating a substantially weaker superconducting-related THG signal compared with the batch~\#1 sample. Correlating the electrical transport data with the extracted THG intensity (Fig.~\ref{S2_0.42}(d)), a residual THG signal is observed above the superconducting onset temperature; nevertheless, due to its low amplitude, it merges with the noise floor above 50~K. These results underscore the sample-dependent nature of the ordered-phase formation. Higher-quality samples favor the stabilization of these ordered phases, which is reflected in a more pronounced terahertz THG response.

\end{document}